\documentclass{aa}
\usepackage{graphicx}
\usepackage{txfonts}
\usepackage{rotating}
\usepackage{lscape}

\begin{document}
   \title{Detecting emission lines with XMM-Newton in 4U 1538$-$52}

   \author{J. J. Rodes-Roca\inst{1,2,3}, K. L. Page\inst{3}, J. M. Torrej\'on\inst{1,2},
          J. P. Osborne\inst{3}
          \and
          G. Bernab\'eu\inst{1,2}
          }

   \offprints{J. J. Rodes-Roca}

   \institute{Department of Physics, Systems Engineering and Sign Theory, University of Alicante,
              03080 Alicante, Spain\\
              \email{rodes@dfists.ua.es}\\
         \and
              University Institute of Physics Applied to Sciences and Technologies,
              University of Alicante, 03080 Alicante, Spain\\
         \and
             Department of Physics and Astronomy, University of Leicester, Leicester,
             LE1 7RH, UK\\
                          }

   \date{Received     ; accepted       }

  \abstract
   {The properties of the X-ray emission lines are a fundamental tool for
   studying the nature of the matter surrounding the neutron star and
   the phenomena that produce these lines.}
   {The aim of this work is to analyze the X-ray spectrum of 4U 1538$-$52
   obtained by the \emph{XMM-Newton} observatory and to look for the presence
   of diagnostic lines in the energy range 0.3--11.5 keV.}
   {We used a 54 ks \emph{PN \& MOS/XMM-Newton} observation of the high mass X-ray binary
   4U 1538$-$52 covering the orbital phase between 0.75 to 1.00 (the eclipse-ingress).
   We have modelled the 0.3--11.5 keV continuum emission
   with three absorbed power laws and looked for the emission lines.}
   {We found previously unreported recombination lines, in this system, at $\sim$2.4 keV,
   $\sim$1.9 keV and $\sim$1.3 keV, consistent with the presence of highly ionized
   states of S \textsc{XV} He$\alpha$, Si \textsc{XIII} He$\alpha$ and Mg K$\alpha$ or
   Mg \textsc{XI} He$\alpha$. On the other hand, both out of eclipse and in eclipse
   we detect a fluorescence iron emission line at 6.4 keV which is resolved
   into two components: a narrow ($\sigma \leq 10$ eV) fluorescence Fe
   K$\alpha$ line plus one hot line from highly photoionized Fe \textsc{XXV}.}
   {The detection of new recombination lines during eclipse-ingress in 4U 1538$-$52
   indicates that
   there is an extended ionized region surrounding the neutron star.}

   
   \authorrunning{J. J. Rodes-Roca et al.}
   \titlerunning{Detecting emission lines in 4U 1538$-$52}
   \maketitle
%

\section{Introduction}

4U 1538$-$52, discovered by the \emph{Uhuru} satellite (\cite{giacconi74}),
is an X-ray pulsar with a B-type supergiant companion,
QV Nor.
It has an orbital period of $\sim$3.728 days (\cite{DWP77}; \cite{clark00}),
with eclipses lasting $\sim$0.6 days (\cite{becker}).
This X-ray persistent system produces this radiation when the neutron star
captures matter from the wind of the B supergiant star. Assuming a distance
to the source of 5.5 kpc (\cite{becker}) and isotropic emission,
the estimated X-ray luminosity is $\sim (2-7)\times 10^{36}$ erg s$^{-1}$
in the 3$-$100 keV range (\cite{rodesPhD}).
Therefore, the size of the ionization zone may be
a relatively small region in the stellar wind
(\cite{HM77}; \cite{vLoon01}).

Fluorescence iron emission lines from X-ray pulsars are produced
by illumination of neutral or partially ionized material by X-ray
photons with energies above the line excitation energy.
Possible
sites of fluorescence emission may be:
(i) accretion disk (mostly seen in low mass X-ray binary
sources); (ii) stellar wind (in high mass X-ray binary pulsars);
(iii) material in the form of a circumstellar shell; (iv) accretion column;
(v) material in the line of sight; or some combination of these locations.
In this sense, fluorescence lines in the X-ray spectrum are
an interesting tool for studying the surrounding wind regions and
elemental abundance in X-ray sources.

In 4U 1538$-$52 the emission line at $\sim$6.4 keV can usually be described
within the uncertainties either by a single narrow Gaussian line
or by a multiplet of narrow Gaussian lines (\cite{WSH83}).
Observations carried out with \emph{Tenma} detected
it at 6.3$\pm$0.2 keV and EW 50$\pm$30 eV (\cite{makishima87} 1987),
while the Rossi X-ray Timing Explorer (\emph{RXTE}) saw it
at 6.25$\pm$0.06 keV and EW 61 eV (\cite{coburn2}).
Other X-ray observatories such as \emph{BeppoSAX} (\cite{robba})
and \emph{RXTE} have used a Gaussian line
at $\sim$6.4 keV for describing the fluorescence of iron in a
low-ionization region (\cite{mukherjee06} 2006; \cite{rodesPhD}). The variability of
this line was studied by \cite{jjrrXrU05}.

In this paper, we present a spectral analysis based on the
observation of 4U 1538$-$52 performed with data from the \emph{XMM-Newton}
satellite. The observation covers the orbital phase interval
0.75--1.00 and we detect the presence of the K$_\alpha$ iron
line at $\sim$6.4 keV and some blended emission lines
below 3 keV. In Sect.~\ref{data} we describe the
observation and data analysis. We present in Sect.~\ref{timing}
timing analysis; in Sect.~\ref{analyse} spectral
analysis, and in Sect.~\ref{conclusion} we summarize our results.

\section{Observation}
\label{data}

4U 1538$-$52 was observed with the \emph{XMM-Newton} satellite in
2003 from August 14 15:34:01 to August 15 14:02:30 UT using \emph{EPIC/PN} and
\emph{EPIC/MOS}. The three European Photon Imaging Cameras (\emph{EPIC})
consist of one \emph{PN}-type CCD camera and two \emph{MOS} CCD cameras
(\cite{struder}; \cite{turner}). The three \emph{EPIC} cameras were operated
in \emph{Full frame} mode, with a time resolution of 73.4 ms for the \emph{PN}
camera and of 2.6 s for the two \emph{MOS} cameras, and the thin filter $1$
(\emph{MOS-1} and \emph{PN}) and medium filter (\emph{MOS-2}) were used.
The net count rate of each one was
$3.116\pm0.008$ cts s$^{-1}$, $0.723\pm0.003$ cts s$^{-1}$ and
$0.734\pm0.003$ cts s$^{-1}$, respectively.
The reflection grating spectrometer (\emph{RGS}) showed a low level
of counts, $1.4\pm1.1 \times 10^{-3}$ cts s$^{-1}$, and we did not use
this spectrum in our analysis.
The resulting effective exposure time was 53.68 ks for the \emph{PN},
64.33 ks for the \emph{MOS-1} and 67.93 ks for the \emph{MOS-2}.
The details of the observation are listed in Table\ref{obsdata}.
To calculate the orbital phase we used the best-fit ephemeris
of \cite{makishima87} (1987) with the orbital period $P_{orb} = 3.72854\pm0.00015$
days and the eclipse centre 45,518.14 days (Modified Julian Date).

\begin{table}
  \caption{Log of \emph{XMM-Newton} observation of 4U 1538$-$52.
  }
\label{obsdata}
\centering                          
\begin{tabular}{l c c c}        
\hline\hline                 
Instrument & MJD & Exposure &  Orbital  \\    
 & Start--End & (ks) & phase  \\
\hline                        
{\rm PN} & 52 865.66--52 866.58 & 53.68 & 0.75--1.00 \\      
{\rm MOS-1} & 52 865.65--52 866.58 & 64.33 & 0.75--1.00 \\
{\rm MOS-2} & 52 865.65--52 866.58 & 67.93 & 0.75--1.00 \\      
\hline                                   
\end{tabular}
\end{table}

We reduced the data using Science Analysis System (SAS) version 8.0, using
the most up-to-date calibration files. Spectra and light curves were
extracted using circular regions centered on the source position, with
radii of 600 pixels (30 arcsec) for \emph{EPIC/PN} and 400 pixels (20 arcsec)
for the two \emph{EPIC/MOS} instruments. We selected background data from
circular regions offset from the source, with radii six times the source extraction
radius. We accumulated all the events with patterns 0--4 and 0--12 for
\emph{EPIC/PN} and \emph{EPIC/MOS}, respectively.

As 4U 1538$-$52 is a bright source, we analyzed whether our data were affected by pile-up,
i.e. if more than one X-ray photon arrives at either a single CCD pixel, or adjacent pixels,
before the charge is read out. This effect can distort the source spectrum and the measured flux.
The XMMSAS provides the task EPATPLOT to determine the distribution patterns
produced on the CCDs by the incoming photons and verify whether the source
is piled-up or not. We created an event list extracted for both an annulus
and a circle region and displayed the fraction of single, double, triple and quadruple
pixel events.
The resulting plots of the \emph{PN} and \emph{MOS} data revealed no obvious pile-up,
even when using a circle.
The maximum count rate before pile-up is typically
$\sim$0.7 cps for the \emph{MOS} instruments and 6 cps for \emph{PN} in full frame mode.
Therefore, we conclude that our data are not affected by pile-up.

\section{Timing analysis}
\label{timing}

\begin{figure*}[htb]
  \centering
  \includegraphics[angle=-90,width=\textwidth]{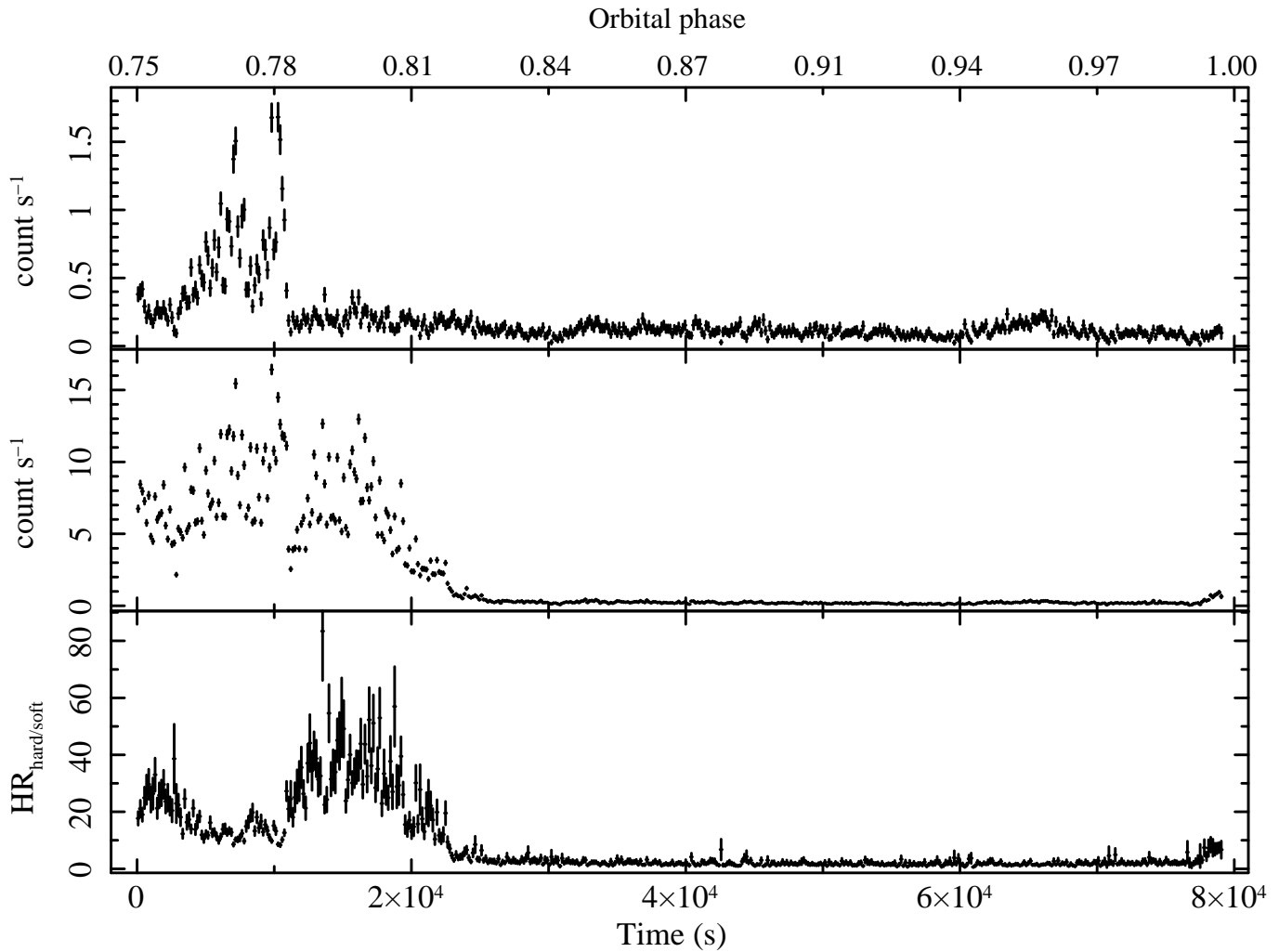}
  \caption{\emph{EPIC/PN} light curve of 4U 1538$-$52,
  binned at 20 s, in the energy ranges 0.3--3 keV (top panel)
  and 3--11.5 keV (middle panel), and their hardness ratio (bottom panel). 
  }
  \label{lcurve1}
\end{figure*}

In Fig.~\ref{lcurve1} we show the background subtracted light curves
for the \emph{EPIC/PN} camera, binned at 20 s, in the energy ranges 0.3--3 keV
and 3--11.5 keV and their hardness--ratio (HR) calculated as (3--11.5 keV)/(0.3--3 keV)
(hard/soft).
This light curve shows the source out of eclipse during the first $\sim$25 ks
and in eclipse in the last $\sim$55 ks. The shape of the light curve is
similar both in the hard and soft components during
the eclipse. During the first $\sim$10 ks there is a softening trend
but the HR presents significant differences between $\sim$10 ks and
$\sim$20 ks. In this time interval, the soft component has a count rate similar
to that in eclipse, whereas the hard component appears uneclipsed.

\begin{figure}[htb]
  \centering
  \includegraphics[angle=-90,width=9cm]{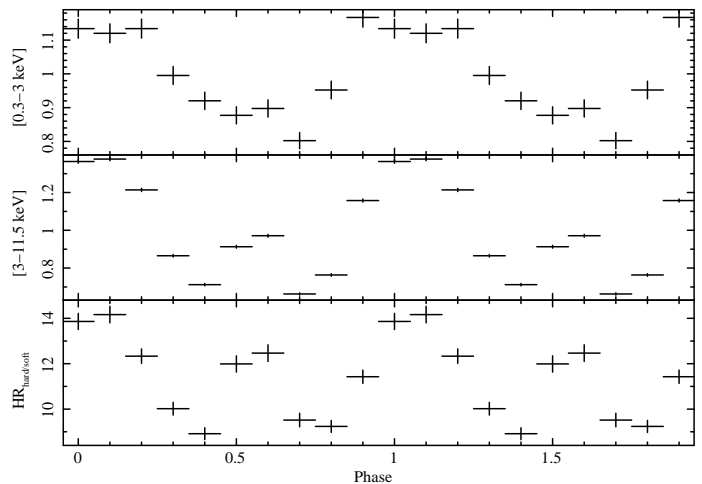}
  \caption{\emph{EPIC/PN} folded light curves of 4U 1538$-$52,
  binned at 20 s, in the energy ranges 0.3--3 keV (top panel),
  3--11.5 keV (middle panel) and their hardness ratio
  (3--11.5)/(0.3--3) (bottom panel).
  }
  \label{HR_PN}
\end{figure}

The \emph{XMM-Newton} observation of 4U 1538$-$52 allows us to investigate the
pulse profile in detail. To estimate the pulse period, we used the \emph{XRONOS} version 5.21
timing analysis software package. First of all, we estimated the pulse period with
the \emph{powspec} task; secondly, we searched for the best pulse period with the \emph{efsearch}
task using the $\chi^2$ test; and finally, we obtained the folded light curve at the best-fit
period $P = 526.7 \pm 0.2$ s. This result is consistent with the \emph{RXTE} pulse period
obtained by \cite{mukherjee06} (2006), $P = 526.849 \pm 0.003$ s.
Fig.~\ref{HR_PN} shows the folded
light curves in the two energy intervals 0.3--3 keV (soft) and 3--11.5 keV (hard)
together with the folded hardness--ratio. The HR
shows two peaks at pulse phases 0.1 and 0.6, respectively.

\section{Spectral analysis}
\label{analyse}

For spectral analysis we used the \textsc{XSPEC} version 12.5.0 (\cite{arnaud}) fitting package,
released as a part of \textsc{XANADU} in the HEASoft tools.

The \emph{EPIC} cameras cover the energy range 0.3--11.5 keV. In previous
works based on a wider energy range (\emph{Tenma}, 1.5--35 keV, \cite{makishima87} 1987;
\emph{BeppoSAX}, 0.1--100 keV, \cite{robba}; \emph{RXTE}, 3--100 keV,
\cite{coburn2}; \cite{rodesPhD}; \emph{INTEGRAL}, 3--100 keV, \cite{jjrr09})
the X-ray continuum of 4U 1538$-$52 was described by
different absorbed power law relations, modified by a high-energy cutoff
plus a Gaussian emission line at $\sim$6.4 keV, whenever present. Moreover,
these models have also been modified by absorption cyclotron resonant scattering
features at $\sim$21 keV (\cite{clark90}; \cite{robba}) and at
$\sim$47 keV (\cite{jjrr09}).

Other X-ray continuum models of high-mass X-ray binaries (HMXBs),
such as two absorbed power laws or three
absorbed power laws, can also describe the observations with \emph{ASCA} of
Vela X--1 (\cite{sako99}) or with \emph{Chandra} \& \emph{XMM-Newton} of 4U 1700$-$37
(\cite{boroson03}; \cite{vdMeer05} 2005). One power law is used to fit the direct
continuum which originates near the neutron star. Part of this radiation is
scattered by the stellar wind of the supergiant star and is fitted by
the second power law. Some HMXBs present a soft excess
at low energies (0.1--1 keV) which may be modelled as another absorbed power
law, blackbody component or bremsstrahlung component. The physical origin of this component
is not well understood, although it is reported for some X-ray binary sources,
such as 4U 1700$-$37 (\cite{vdMeer05} 2005), 4U 1626$-$67, Cen X-3 or Vela X-1
(\cite{HNK04}). The \emph{BeppoSAX} spectrum of 4U 1538$-$52 showed a soft excess
at 0.3--1.0 keV, which was modelled with a blackbody component with a temperature
of 0.08 keV (\cite{robba}). Moreover, they also obtained a good fit using a bremsstrahlung
component with a temperature of 0.10 keV. We used all of these models
to fit the \emph{XMM-Newton} continuum spectrum of 4U 1538$-$52 and Gaussian functions
to describe the emission lines. 

\subsection{X-ray continuum}
\label{continuum}

\textbf{In order to look for diagnostic emission lines a proper description
of the underlying continuum is critical. Because of its higher count rate
we used only the \emph{PN} spectrum to pin down the parameters of the
phase averaged spectrum and then we combined both \emph{PN} and
\emph{MOS} pulse phase averaged spectra in order to check our results.
}
We fitted the 0.3--11.5 keV energy spectrum using three absorbed power-law
components \textbf{fixing the photon index parameter to the same value for each
power law component}, but with different normalization
and column density $N_H$. We modelled the hard X-ray continuum (3--11.5 keV)
using two power laws, each of them modified by different absorbing columns
(\cite{MMcC83}). We obtained adequate fits using the same photon index
$\alpha\sim 1.134^{+0.018}_{-0.023}$ for both components which describe the X-ray continuum radiation
from the neutron star: a direct emission from the compact object, absorbed
through the surrounding stellar wind, and scattered radiation produced by
Thomson scattering by electrons in the extended stellar wind of the
supergiant star.
\textbf{Taking the hydrogen columns listed in Table~\ref{xmm-ncont}
into account, the hard component is assumed to originate from the
near surroundings of the neutron star and is highly absorbed by optically
thick matter in the line of sight ($N_H \sim 6.0\times10^{23}$ cm$^{-2}$),
and the scattered component is produced by optically thin plasma
($N_H \sim 1.2\times10^{23}$ cm$^{-2}$). Other HMXBs with spectra which
can be described by these components are Cen X--3 (\cite{nagase92} 1992;
\cite{ebisawa96}) and 4U 1700--37 (\cite{vdMeer05} 2005). These systems
also showed a hard component highly absorbed even during the eclipse.
}

\begin{figure}[htb]
  \centering
  \includegraphics[angle=-90,width=9cm]{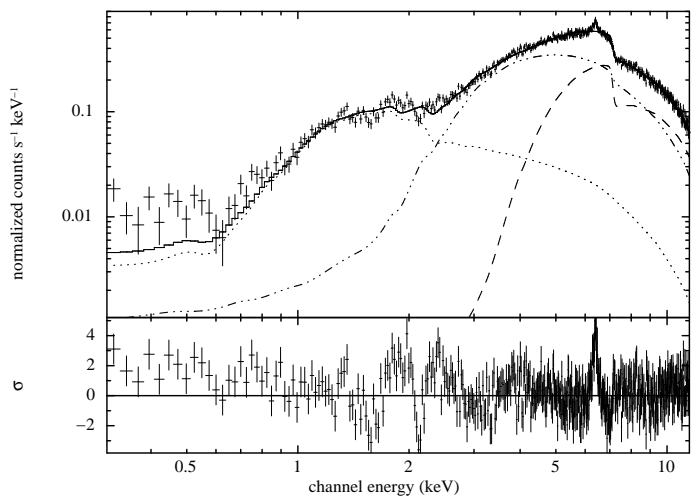}
  \caption{\emph{EPIC/PN} spectrum of 4U 1538$-$52 and X-ray continuum model.
  The dash line represents the direct component, the dash-dot-dot-dot the scattered
  component, and the dot line the soft component (top panel). The lower panel
  shows the residuals between the spectrum and the model.
  }
  \label{PNcontinuum}
\end{figure}

We used the third absorbed power-law because a soft excess at 0.3--1.0
keV is detected in the \emph{PN} spectrum. We fixed the photon index to
that of the other power-laws obtaining a significant improvement
($\chi^2_\nu = 1.16$, with an F-test of 4.2$\times 10^{-131}$).
Fig.~\ref{PNcontinuum} shows the three different absorbed power-law
continuum components and their residuals in units of $\sigma$.
The spectrum clearly shows several emission lines at 6.4 keV,
2.4 keV, 1.9 keV and 1.3 keV, and an absorption feature at 2.1 keV.
The flux ratios are usually above 30\%.
We are sure that the low energy residuals are not due to incomplete
calibration. The cross-calibration \emph{XMM-Newton} database consists of
$\sim$150 observation of different sources, optimally reduced, fit with
spectral models defined on a source-by-source
basis\footnote{http://www.iachec.org/meetings/2009/Guainazzi\_2.pdf}.
As a result, deviations for \emph{PN} and \emph{MOS} flux ratios above
10\% are consistent with the presence of emission or absorption lines.

\begin{table}
\begin{minipage}[t]{\columnwidth}
  \caption{Fitted parameters for the \emph{PN} X-ray continuum 
  spectrum in Fig.~\ref{PNcontinuum}.
  }
\label{xmm-ncont}
\centering                          
\renewcommand{\footnoterule}{}      
\begin{tabular}{r r c c}        
\hline\hline                 
Component & Parameter & 3PL\footnote {Three absorbed power-laws} & 2PL+bb\footnote{Two absorbed power-laws plus an absorbed blackbody} \\    
\hline                        
{\rm Hard} & $N_H$ (10$^{22}$ cm$^{-2}$) & 60 $\pm$ 4 & 62$^{+4}_{-5}$ \\      
  & Photon index\footnote{The same for all the power-laws} & 1.134$^{+0.018}_{-0.023}$ & 1.13$^{+0.03}_{-0.05}$ \\
  & Normalization\footnote{($\times$ 10$^{-4}$) in units of photons s$^{-1}$ cm$^{-2}$ keV$^{-1}$ at 1 keV} & 93 $\pm$ 9 & 92$^{+11}_{-10}$ \\
  &  &  & \\
{\rm Scattered}  & $N_H$ (10$^{22}$ cm$^{-2}$) & 11.70$^{+0.21}_{-0.4}$ & 12.7$^{+0.7}_{-0.9}$ \\
  & Normalization$^d$ & 43 $\pm$ 3 & 47$^{+4}_{-3}$ \\
  &  &  & \\
{\rm  Soft} & $N_H$ (10$^{22}$ cm$^{-2}$) & 0.74 $\pm$ 0.04 & 0.42$^{+0.04}_{-0.03}$ \\
  & Normalization$^d$ & 2.42$^{+0.15}_{-0.14}$ & $\cdots$ \\
  & $kT_{bb}$ (keV) & $\cdots$ & 0.925$^{+0.015}_{-0.016}$ \\
  & Normalization$_{bb}$ & $\cdots$ & 1.8$^{+1.1}_{-0.5}\times 10^{-5}$ \\
 & & & \\
 &  $\chi^2_{\nu}$(dof) & 1.16(2235) & 1.16(2234) \\ 
\hline                                   
\end{tabular}
\end{minipage}
\end{table}

We also used other components to investigate the physical origin of this soft-excess, such as
a bremsstrahlung or blackbody component.
These model components need to be modified by an
absorption column too. Although a bremsstrahlung component improves the fit significantly
($\chi^2_\nu = 1.17$, F-test of 4.0$\times 10^{-131}$), its inferred temperature is too high
(196.9 keV) and could not be constrained within the \emph{XMM} band pass,
suggesting that the soft excess has another physical origin.

In Table~\ref{xmm-ncont} we report the model
parameters of the X-ray continuum spectrum for the \emph{PN} camera.
All uncertainties refer to a single parameter at the 90\% ($\Delta\chi^2 = \; 2.71$)
confidence limit\footnote{Also in Tables \ref{PNMOSemli}, \ref{pulsefase},
\ref{eclcont} and \ref{emlieclparam}}.

When we used a blackbody component we also obtained a similar improvement
of the fit ($\chi^2_\nu = 1.16$, F-test of 1.6$\times 10^{-129}$). The blackbody has a temperature of
$kT\sim 0.925^{+0.015}_{-0.016}$ keV. We calculated the luminosity of
the source of the blackbody component over the 0.3--11.5 keV energy band, assuming a distance
to the source of 5.5 kpc (\cite{becker}) and isotropic emission
to be 5.4$\times 10^{33}$ erg s$^{-1}$. Since the surface luminosity of a blackbody
depends only on its temperature, it is possible to calculate the size of the
emitting region:
   \begin{equation}\label{radius}
   R_{bb} (km) = 3.04 \times 10^4 \frac{D \, \sqrt{F_{bb}}}{T^2_{bb}} \,,
   \end{equation}
where $D$ is the distance to the source in kpc, $F_{bb}$ the unabsorbed flux
over the 0.3--11.5 keV energy band in
erg s$^{-1}$ cm$^{-2}$, $T_{bb}$ the temperature in keV and $R_{bb}$ is the
radius of the emitting region. From Eq.\ref{radius}, we found a radius of
the emitting surface of 0.24 km. If we assume thermal emission from the
neutron star polar cap, this radius may be consistent with the
expected size. However, we estimated
the radius of the accreting polar cap using the equation (\cite{HNK04}):
   \begin{equation}\label{radius2}
   R_{col} = R_{NS} \sqrt{\frac{R_{NS}}{R_m}} \,,
   \end{equation}
where $R_m$ is the magnetospheric radius and $R_{NS}$ the neutron star radius.
The magnetospheric radius may be estimated from the magnetic dipole moment and mass
accretion rate (\cite{EL1977}; \cite{AKB96}):
   \begin{equation}\label{alfven}
   R_{m} (cm) = 3.2 \times 10^8 \, \, \dot{M}^{-2/7}_{17} \; \;
                {\mu}^{4/7}_{30} \, \, \left(\frac{M}{M_{\odot}}\right)^{-1/7} \,,
   \end{equation}
where $\mu_{30}$ is the magnetic dipole moment of the neutron star in units of
10$^{30}$ G cm$^3$ and $\dot{M}_{17}$ is the mass accretion rate in units of
10$^{17}$ g s$^{-1}$. With $M = 1.3 M_{\odot}$ (\cite{reynolds92}),
$\dot{M}_{17} = 883$ g s$^{-1}$ (\cite{clark94}; \cite{rodesPhD}; \cite{rodes2008})
and $\mu_{30} = 1.8$ G cm$^3$ (\cite{rodesPhD}), the Eq.~\ref{alfven} yields the
value $R_m \sim$ 6.2$\times 10^{7}$ cm.
Assuming $R_{NS}$ $\sim$ 10$^6$ cm, we obtained $R_{col} \sim$ 1.27 km,
five times our estimated blackbody emitting radius. \textbf{Although this
result suggests that the soft excess is not formed by blackbody emission,
taking into account the associated errors in the parameters, we can not exclude
blackbody emission as the origin of the soft excess}. 

We also tried to fit this soft excess by adding 2 to the photon index of the soft component
compared to the index of the hard component, i.e. scattering by dust grains at large
distance from the source (\cite{robba}; \cite{vdMeer05} 2005).
We obtained a good description of the spectrum
($\chi^2_\nu = 1.20$, F-test of 11$\times 10^{-116}$), but a significantly worse fit
than the previous components ($\chi^2_\nu = 1.16$).

Finally, we modelled the soft X-ray continuum component with a blend of Gaussians only
(\cite{boroson03}; \cite{vdMeer05} 2005) and refitted the data. We could not obtain
a good description below 0.6 keV, though. \textbf{Therefore, although the model composed
of three power-law components also does not fit the data well below 0.6 keV, it was
the best model obtained and we used it for our analysis.}

\textbf{Afterwards we combined \emph{EPIC/PN} and \emph{EPIC/MOS} spectra
and applied the same models to them obtaining similar results.
The spectra of all three \emph{EPIC} instruments were fitted simultaneously, including
a factor to allow for the adjustment of efficiencies between different instruments.
We fixed the continuum parameters, described in Table~\ref{xmm-ncont}, and looked for
the emission features (see Sect.~\ref{fluorescence}).}

\begin{figure}[h!t]
  \centering
  \includegraphics[angle=-90,width=9cm]{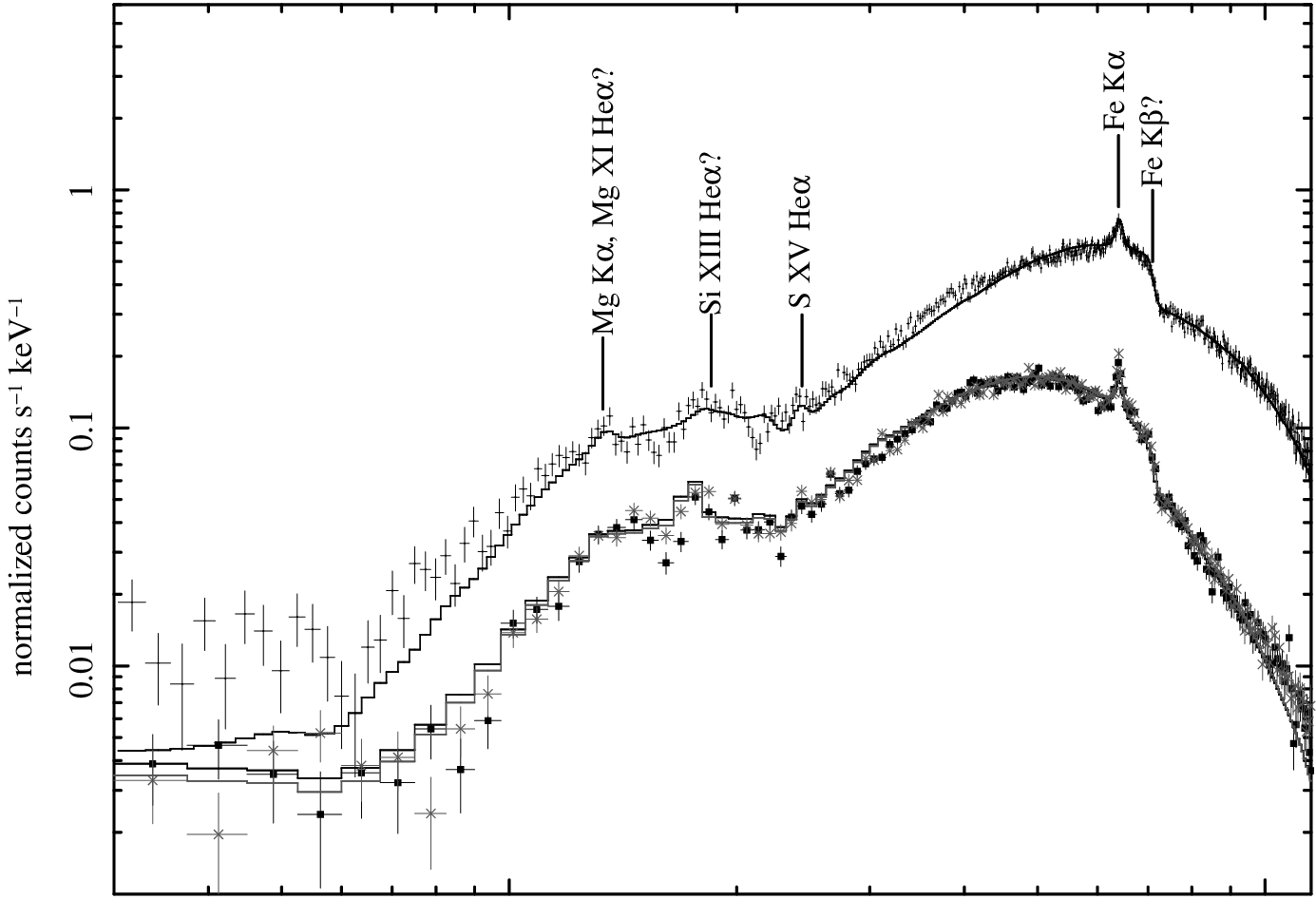}
  \includegraphics[angle=-90,width=9cm]{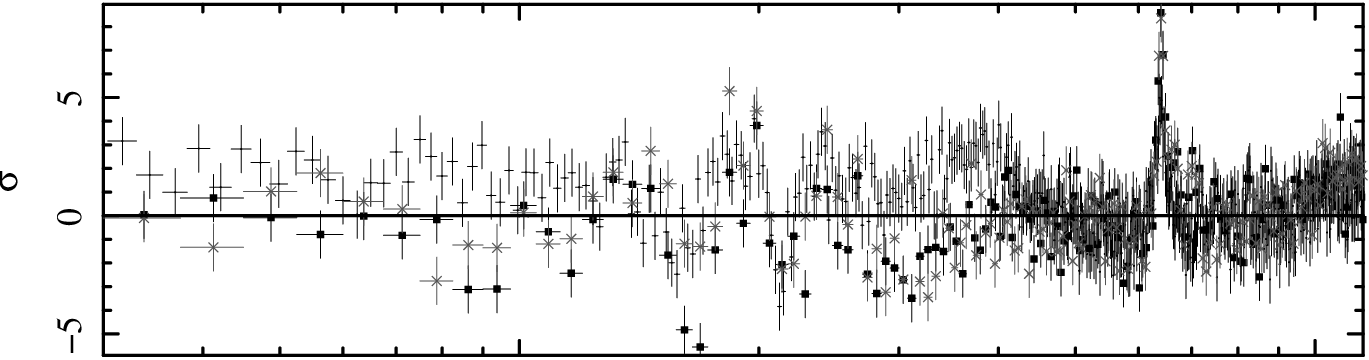}
  \includegraphics[angle=-90,width=9cm]{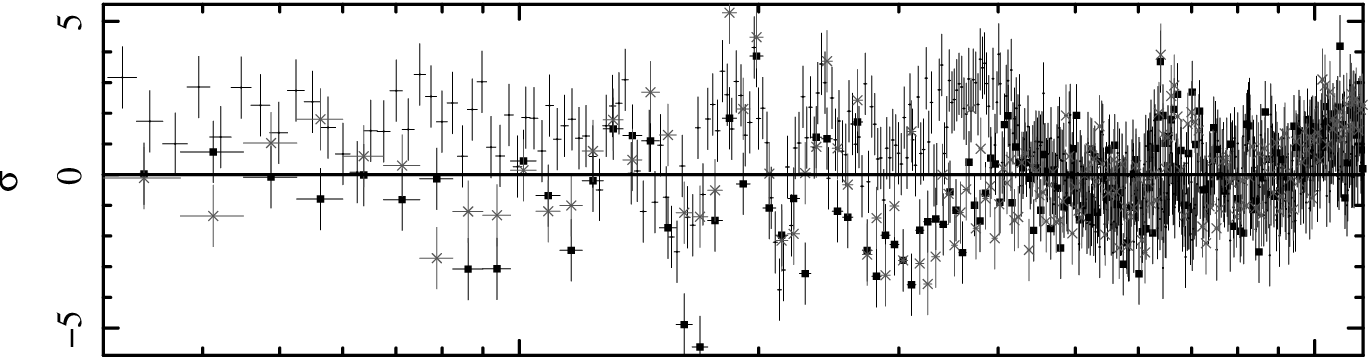}
  \includegraphics[angle=-90,width=9cm]{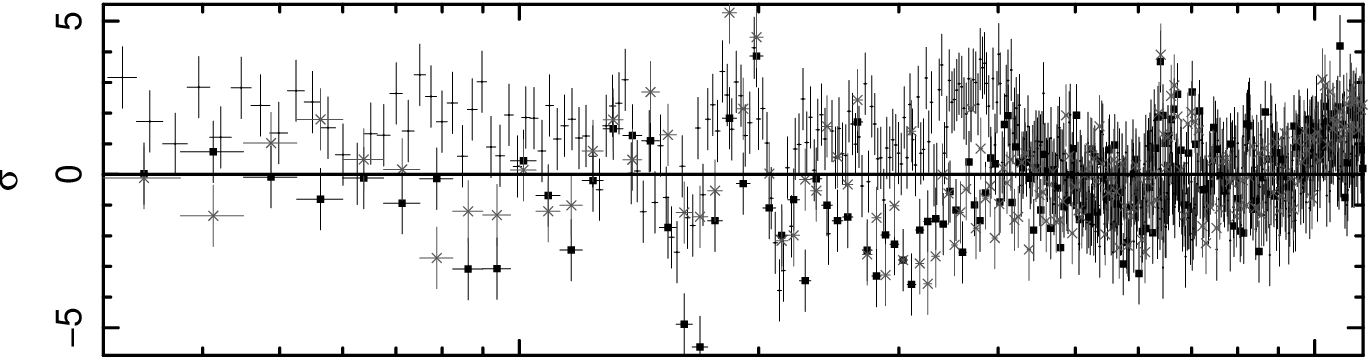}
  \includegraphics[angle=-90,width=9cm]{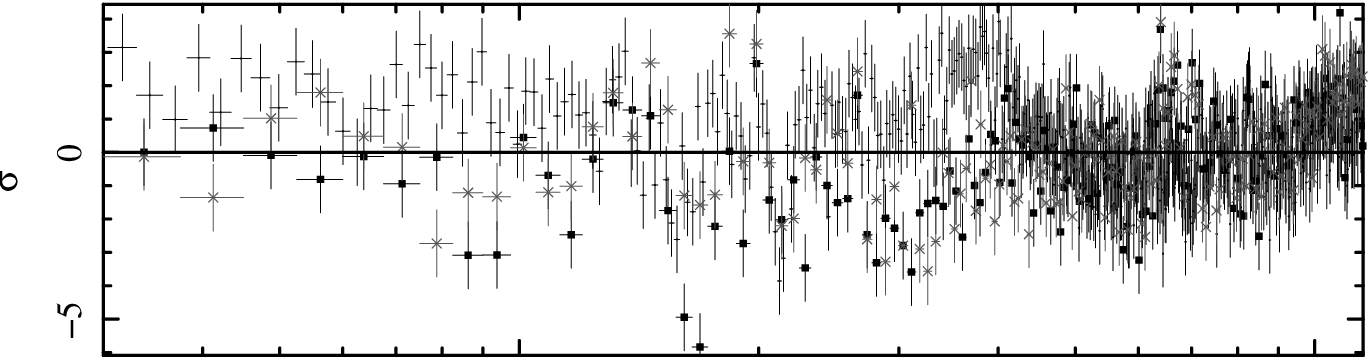}
  \includegraphics[angle=-90,width=9cm]{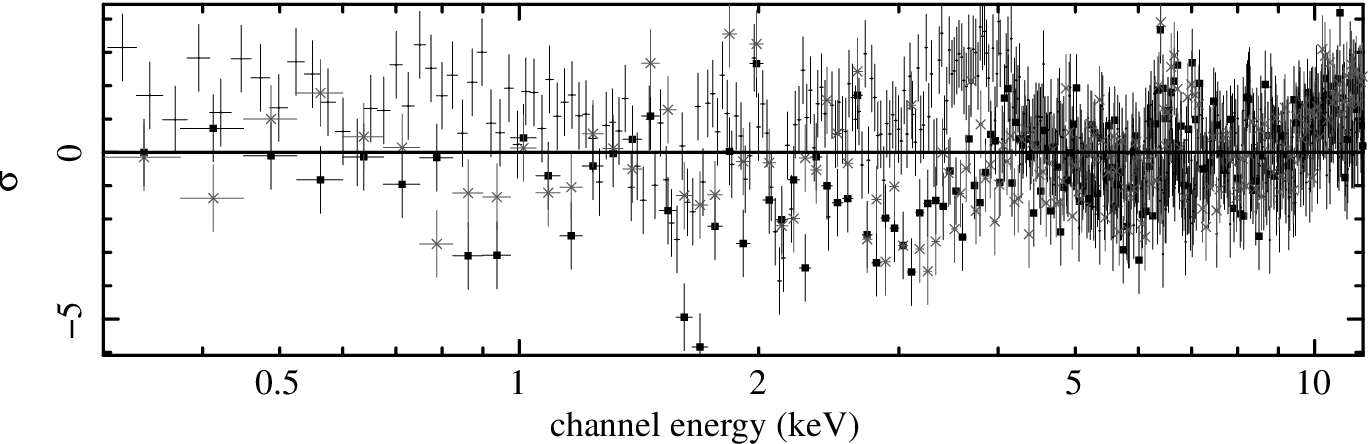}
  \caption{Fluorescence emission lines modeling of the phase-averaged spectra
    \emph{PN} \& \emph{MOS} of 4U 1538$-$52.
    \emph{Top panel}: Spectra and best fit model (three absorbed power-laws and
    four emission lines)
    obtained with \emph{EPIC/PN} and
    \emph{EPIC/MOS}. Bottom panels show the residuals in units of
    $\sigma$ for models taking different numbers of emission lines into account
   (MOS-1 black filled square and MOS-2 grey cross).
    \emph{Second panel}: Without emission lines.
    \emph{Third panel}: Fluorescence iron emission line at $\sim$6.4 keV.
    \emph{Fourth panel}: Fluorescence iron emission line plus a recombination line
    at $\sim$2.42 keV.
    \emph{Fifth panel}: Fluorescence iron emission line plus two recombination lines
    at $\sim$2.42 keV and $\sim$1.90 keV.
    \emph{Bottom panel}: Fluorescence iron emission line plus three recombination lines
    at $\sim$2.42 keV, $\sim$1.90 keV and $\sim$1.34 keV.
    }
  \label{PNMOS}
\end{figure}

\subsection{Emission lines}
\label{fluorescence}

The X-ray eclipse spectra of some HMXBs show many emission lines and radiative
recombination continua (e.g. Vela X-1, \cite{sako99}; Cen X-3 \cite{ebisawa96};
4U 1700$-$37 \cite{vdMeer05} 2005). The data together with the best-fit model,
three absorbed power laws plus emission lines,
and residuals of the fit as the difference between observed flux and model
flux divided by the uncertainty of the observed flux, i.e. in units of
$\sigma$, are included in Fig.~\ref{PNMOS}.
As we can see in Fig.~\ref{PNMOS}, the \emph{PN} spectrum of 4U 1538$-$52
shows emission lines below 3 keV and between 6 and
7 keV. We fitted these lines as Gaussian profiles (see Table~\ref{PNMOSemli} for
parameter values). Because the energy resolution of the \emph{EPIC} cameras is not
sufficient to resolve all lines (FWHM 80--100 eV in the energy range 1--3 keV),
many of them could be blended (e.g., 1.75 keV Si K$\alpha$, Al \textsc{xiii} Ly$\alpha$
and 1.85 keV Si \textsc{xiii} He$\alpha$).
Using the list of emission lines in \cite{vdMeer05}
(2005), we could identify suitable candidates among
fluorescence emission lines from near-neutral
species and discrete recombination lines from He- and H-like species
(see also \cite{drake88} and \cite{hojnacki07}).

Fig.~\ref{PNMOS} clearly shows the presence of the Fe K$\alpha$ line at
$\sim$6.4 keV, but a second broad emission line around 6.65 keV
or an absorption edge at $\sim$7.2 keV associated with low-ionization
iron is an unresolved issue within the deviation for \emph{PN} and \emph{MOS}
ratios.

We started by modeling the fluorescence iron line (see Fig.~\ref{PNMOS}) that was also
detected by several other satellite observatories, such as \emph{Tenma}, \emph{BeppoSAX} or
\emph{RXTE}. Starting with the 3PL continuum model,
after the
inclusion of the emission line at $E\sim 6.4$ keV, the
$\chi^2_{\nu}$ improves from 1.16 for 2242 degrees of freedom (dof) to
1.08, for 2239 dof (F-test of $9.2\times 10^{-35}$).
Although residuals improve significantly above 3 keV
(see Fig.~\ref{PNMOS} third panel),
other emission features can be seen at $\sim$2.4, $\sim$1.9 keV and $\sim$1.3 keV.
Including an emission line at $\sim$2.4 keV improves the fit further resulting in a
$\chi^2_{\nu}$ of 1.07 for 2236 dof (F-test of $9.5\times 10^{-6}$,
see Fig.~\ref{PNMOS} fourth panel). Adding another one at $\sim$1.9 keV,
we obtained a $\chi^2_{\nu}$ of 1.06 for 2233 dof (F-test of $2.6\times 10^{-6}$,
see Fig.~\ref{PNMOS} fifth panel). Finally, the last emission line at
$\sim$1.3 keV marginally improves the fit, leading to $\chi^2_{\nu} =$ 1.05
for 2230 dof (F-test of $0.4$, see Fig.~\ref{PNMOS} bottom panel).

\textbf{Using the combined spectra,
we did not obtain an improvement in the fit quality
($\chi^2_\nu = 1.27$, compared to $\chi^2_\nu = 1.16$ obtained using only the \emph{PN}
spectrum).
Nevertheless, as evident from Fig.\ref{PNMOS}, the
emission lines are clearly seen in the raw data of the three cameras.
In Table~\ref{PNMOSemli} we show the best fit parameters for the emission lines we
detected in Fig.~\ref{PNMOS}, where the top spectrum in the top panel is from
\emph{PN} and the bottom spectra in the top panel are from \emph{MOS-1} (black filled square)
and \emph{MOS-2} (grey cross), respectively.
}

\begin{table}
  \caption{Fitted parameters for the emission lines detected
  in Fig.~\ref{PNMOS} using \emph{PN} and \emph{MOS} data simultaneously.
    }
\label{PNMOSemli}
\centering                          
\begin{tabular}{r r c c}        
\hline\hline                 
Component & Parameter & 3PL & 2PL+bb   \\    
\hline                        
{\rm Fluorescence} & Energy (keV) & 6.4169$^{+0.0011}_{-0.008}$ & 6.4188$^{+0.0006}_{-0.0008}$ \\      
{\rm iron line} & $\sigma$ (eV) & $\leq$ 23 & $\leq$ 60 \\
 & EW (eV) & 53 $\pm$ 5 & 52 $\pm$ 5 \\
{\rm Identification} & & \multicolumn{2}{c}{Fe K$\alpha$ \textsc{I--XVII}} \\
 & F-test & 9.2$\times$10$^{-35}$ & \\
  &  &  & \\
{\rm Other}  & Energy (keV) & 2.422$^{+0.0023}_{-0.0020}$ & 2.423$^{+0.0023}_{-0.0019}$ \\
{\rm emission lines} & $\sigma$ (eV) & 6$^{+50}_{-6}$ & 8$^{+50}_{-8}$ \\
  & EW (eV) & 26 $\pm$ 8 & 27 $\pm$ 9 \\
{\rm Identification} & & \multicolumn{2}{c}{S \textsc{XV} He$\alpha$?} \\
 & F-test & 9.5$\times$10$^{-6}$ & \\
  &  &  & \\
  & Energy (keV) & 1.905$^{+0.0024}_{-0.03}$ & 1.851$^{+0.0020}_{-0.005}$ \\
  & $\sigma$ (eV) & 68$^{+13}_{-15}$ & $\leq$ 40 \\
  & EW (eV) & 50 $\pm$ 12 & 40$^{+8}_{-12}$ \\
{\rm Identification} & & \multicolumn{2}{c}{Si \textsc{XIII} He$\alpha$?} \\
 & F-test & 2.6$\times$10$^{-6}$ & \\
  &  &  & \\
  & Energy (keV) & 1.342$^{+0.015}_{-0.008}$ & 1.340$^{+0.014}_{-0.012}$ \\
  & $\sigma$ (eV) & $\leq$ 21 & $\leq$ 21 \\
  & EW (eV) & 22$^{+11}_{-8}$ & 27$^{+9}_{-8}$ \\
{\rm Identification} & & \multicolumn{2}{c}{Mg K$\alpha$, Mg \textsc{XI} He$\alpha$?} \\
 & F-test & 0.4 & \\
  &  &  & \\
 &  $\chi^2_{\nu}$(dof) & 1.27(3725) & 1.27(3725) \\ 
\hline                                   
\end{tabular}
\end{table}

The F-test is known to be problematic when used to
test the significance of an additional spectral feature (see
\cite{ftest}), even if systematic uncertainties are not an
issue. However, the low false alarm probabilities may make the
detection of the line stable against even crude mistakes in the
computation of the significance (Kreykenbohm \cite{ingophd}).
Therefore, taking into account these caveats, we can conclude that the
fluorescence emission lines are detected with high significance in the
spectrum of this source and none of these features is the result of
poor calibration.

However, none of these models deals with the strong feature at $\sim$2.1
keV. We added a Gaussian absorption profile at this energy and found
an improvement of the fit, but the high optical depth and its unconstrained
value prevent us from accepting it. We checked the possibility that this
feature could be due to a gain effect by changing the gain and offset values
by up to $\pm$3\% and $\pm$0.05 keV, respectively.
The line profile expressed as ratio of a spectral data to a best-fit
model was always around 70\%, suggesting that it is not a gain effect
and the absorption feature could be real.
\textbf{Moreover, although the feature is less pronounced in a spectrum extracted using
single pixel events only, the \emph{EPIC/PN}
calibration team have confirmed (M. Guainazzi, private communication)
the reality of the absorption feature
at 2.1 keV.
}

\subsection{Pulse phase resolved spectra}
\label{pulsephase}

\textbf{The hardness ratio shown in Fig.~\ref{HR_PN} demonstrates
variation of the spectrum with respect to the spin phase of the neutron
star. We divided the pulse into ten phase intervals of equal length in order
to investigate the X-ray spectrum as function of pulse phase. We produced
energy spectra corresponding to each of these phase intervals, which were
fitted by the same X-ray underlying model used to fit the pulse phase averaged
spectrum, i.e., three absorption power laws plus an iron emission line. The resulting
fit parameters are reported in Table~\ref{pulsefase}.
}
\onecolumn
\begin{landscape}
\begin{table*}
\caption{Results of the fit of the pulse phase resolved spectra in the energy
range 0.3--11.5 keV.}\label{pulsefase}
\centering
\begin{tabular}{lcccccccccc} 
\hline\hline             
 & \multicolumn{10}{c}{Pulse phase} \\ \cline{2-11}
Parameter & 0.05 & 0.15 & 0.25 & 0.35 & 0.45 & 0.55 & 0.65 & 0.75 & 0.85 & 0.95 \\
\hline
{\it Hard component} & \multicolumn{10}{c}{} \\
$N_H$ (10$^{22}$ cm$^{-2}$) & 49 $\pm$ 9 & 67$^{+11}_{-9}$ & 69$^{+12}_{-10}$ & 47$^{+8}_{-6}$ &
67$^{+13}_{-11}$ & 79$^{+13}_{-12}$ & 59$^{+11}_{-10}$ & 48$^{+9}_{-8}$ & 58 $\pm$ 8 &
62$^{+9}_{-8}$ \\
Photon index & 0.93$^{+0.12}_{-0.05}$ & 1.114$^{+0.04}_{-0.013}$ & 1.51$^{+0.09}_{-0.13}$ &
1.65 $\pm$ 0.12 & 1.19 $\pm$ 0.10 & 0.95$^{+0.05}_{-0.09}$ & 0.79$^{+0.03}_{-0.09}$ &
1.16$^{+0.06}_{-0.09}$ & 1.44$^{+0.08}_{-0.11}$ & 1.31 $\pm$ 0.12 \\
Normalization$^a$ &
57$^{+12}_{-11}$ & 90$^{+40}_{-22}$ & 160$^{+60}_{-50}$ & 210$^{+90}_{-50}$ & 103$^{+23}_{-30}$ &
110 $\pm$ 30 & 64$^{+22}_{-17}$ & 112$^{+21}_{-30}$ & 180$^{+90}_{-50}$ & 130$^{+50}_{-40}$ \\
\hline
{\it Scattered component} & \multicolumn{10}{c}{} \\
$N_H$ (10$^{22}$ cm$^{-2}$) & 12.0$^{+2.3}_{-3}$ & 13.6$^{+1.2}_{-1.3}$ & 12.7$^{+1.4}_{-1.3}$ &
9.2$^{+0.8}_{-1.4}$ & 11.4 $\pm$ 1.0 & 12.8 $\pm$ 1.1 & 11.6$^{+0.8}_{-1.5}$ &
10.8$^{+0.7}_{-1.4}$ & 12.0$^{+1.2}_{-1.4}$ & 10.4$^{+1.3}_{-1.2}$ \\
Normalization$^a$ &
21.6$^{+2.1}_{-7}$ & 44$^{+5}_{-6}$ & 54$^{+8}_{-11}$ & 47$^{+18}_{-14}$ & 61$^{+10}_{-9}$ &
49$^{+10}_{-8}$ & 32$^{+7}_{-6}$ & 48$^{+13}_{-11}$ & 59$^{+9}_{-12}$ & 34$^{+10}_{-8}$ \\
\hline
{\it Soft component} & \multicolumn{10}{c}{} \\
$N_H$ (10$^{22}$ cm$^{-2}$) & 0.60$^{+0.14}_{-0.08}$ & 0.76$^{+0.15}_{-0.08}$ & 0.78$^{+0.15}_{-0.11}$ & 0.75$^{+0.12}_{-0.11}$ & 0.67$^{+0.15}_{-0.10}$ & 0.70$^{+0.11}_{-0.09}$ &
0.59$^{+0.10}_{-0.09}$ & 0.69$^{+0.12}_{-0.10}$ & 0.90 $\pm$ 0.12 & 0.67$^{+0.10}_{-0.09}$ \\
Normalization$^a$ &
2.0$^{+0.4}_{-0.18}$ & 2.51$^{+0.16}_{-0.21}$ & 2.8$^{+0.5}_{-0.3}$ & 2.9$^{+0.7}_{-0.6}$ &
2.50$^{+0.4}_{-0.17}$ & 2.5$^{+0.4}_{-0.3}$ & 2.1$^{+0.2}_{-0.3}$ & 2.6$^{+0.5}_{-0.4}$ &
3.6$^{+0.6}_{-0.5}$ & 2.4$^{+0.5}_{-0.4}$ \\
\hline
{\it Iron line} & \multicolumn{10}{c}{} \\
Energy (keV) & 6.390$^{+0.021}_{-0.03}$ & 6.40$^{+0.04}_{-0.03}$ & 6.37 $\pm$ 0.03 &
6.38$^{+0.09}_{-0.10}$ & 6.44 $\pm$ 0.05 & 6.44 $\pm$ 0.11 & 6.39$^{+0.03}_{-0.04}$ &
6.39 $\pm$ 0.03 & 6.41$^{+0.03}_{-0.04}$ & 6.37$^{+0.10}_{-0.09}$ \\
$\sigma$ (eV) & $\leq$ 70 & 30$^{+60}_{-30}$ & $\leq$ 60 & 150 $\pm$ 60 & 60 $\pm$ 60 &
130$^{+100}_{-80}$ & $\leq$ 70 & $\leq$ 50 & 70$^{+40}_{-30}$ & 100$^{+70}_{-80}$ \\
EW (eV) & 54 $\pm$ 16 & 55$^{+20}_{-16}$ & 59$^{+21}_{-16}$ & 70 $\pm$ 30 & 42$^{+18}_{-16}$ &
33$^{+19}_{-30}$ & 35$^{+15}_{-11}$ & 40$^{+15}_{-12}$ & 67$^{+20}_{-19}$ & 60 $\pm$ 30 \\
Unabsorbed flux$^b$ & 1.63 & 2.08 & 1.95 & 2.06 & 2.27 &
3.05 & 2.48 & 2.30 & 2.37 & 1.87 \\
C-statistic(dof) & 2362(2239) & 2289(2239) & 2367(2239) & 2426(2239) & 2337(2239) & 2337(2239) &
2255(2239) & 2302(2239) & 2389(2239) & 2362(2239) \\
\hline
\multicolumn{11}{l}{$^a$($\times$ 10$^{-4}$) in units of photons s$^{-1}$ cm$^{-2}$ keV$^{-1}$ at 1 keV} \\
\multicolumn{11}{l}{$^b$($\times$ 10$^{-10}$) erg s$^{-1}$ cm$^{-2}$} \\
\end{tabular}
\end{table*}
\end{landscape}
\twocolumn

\textbf{We did not find statistically significant residuals in the energy
range between 1 and 3 keV, and therefore no low energy emission lines are required to
fit these spectra, although this could be because of the lower statistics
of the phase resolved spectra with respect to the phase averaged spectrum.
The energy of the iron emission line is compatible with being unchanged along the pulse
profile, although the variations in the depth, width and equivalent width
could be due to the presence of other unresolved emission lines. The
\emph{BeppoSAX} spectrum showed similar results with the pulse phase
of the iron emission line parameters using four phase intervals
(\cite{robba}). In other pulse phase
spectroscopy analysis either the iron emission line was not present
significantly (\emph{EXOSAT}; \cite{robba92}) or was also unchanged
when the pulse phase profile was divided
into two phase intervals (\emph{Tenma}; \cite{makishima87} 1987).
The variation
of the different column absorption parameters are significant
along the pulse profile, but again these do not show a clear relationship
to the other parameters.
}

\begin{figure}[h!t]
  \centering
  \includegraphics[angle=-90,width=9cm]{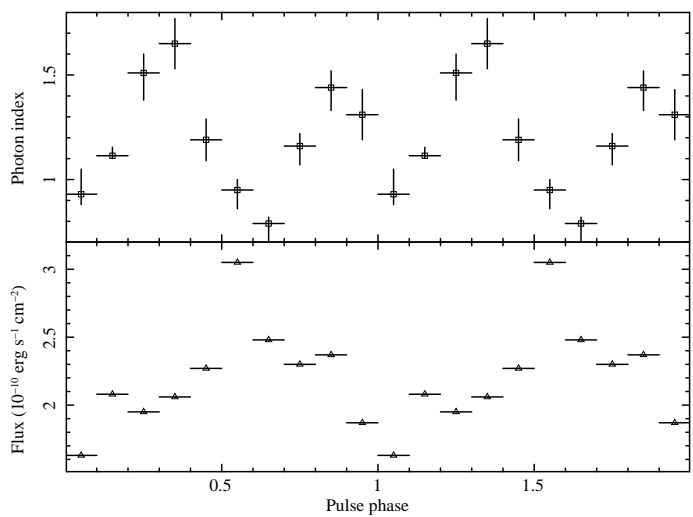}
  \caption{Photon index (top panel) and the corresponding unabsorbed flux in
    the 0.3--11.5 energy band (10$^{-10}$ erg s$^{-1}$ cm$^{-2}$, lower panel)
    versus the corresponding phase interval.
    }
  \label{pf_piflux}
\end{figure}

\textbf{The photon index presents a clear modulation with the pulse phase
showing two maximum values at phases 0.35 and 0.85 and one minimum value
at phase 0.65. In Fig.~\ref{pf_piflux} we plot the photon index (top panel)
versus the corresponding phase interval together with the corresponding
unabsorbed flux in the 0.3--11.5 keV energy band (lower panel). As can be seen from
this figure, the phase of maximum unabsorbed flux is close to the phase of minimum photon
index, however there is not a one-to-one between these two parameters.
The unabsorbed flux presents two flat regions in
the phase intervals 0.15--0.35 and 0.65--0.85 and increasing or decreasing
flux in the other interval phases, it peaks at phase 0.55 and is near the minimum
in the photon index. The result obtained from \emph{BeppoSAX} showed a clear
anticorrelation between the photon index and the flux, with the
spectrum being flatter in correspondence with higher flux levels (\cite{robba}).
Moreover, this behaviour of the photon index from \emph{BeppoSAX} was similar to previous results
from \emph{EXOSAT} (\cite{robba92}) and \emph{Ginga} (\cite{clark90}).
}

\subsection{Detecting emission lines in the eclipse spectrum}
\label{eclipse}

As we showed in Sect.~\ref{timing}, the eclipse of the X-ray source is clearly
visible in the last 55 ks of the \emph{XMM-Newton} observation (see Fig.~\ref{lcurve1}).
Therefore, we analyzed the phase-averaged spectrum out of eclipse and in eclipse.
\textbf{The X-ray continuum spectrum out of the eclipse could be described by the models
used in this work. While the parameters of the fluorescent iron emission line at 6.4 keV
were consistent with those of the whole observation, we did not find evidence of the other
emission lines. Nevertheless, Fig.~\ref{lcurve1} suggests the presence of a
variability out of the eclipse since the HR changes with time. Therefore, we
divided the data out of the eclipse into two temporal intervals, 0--10 ks
and 10--20 ks, taking into account the HR variability. We could describe these
spectra with the same X-ray continuum, i.e. three absorbed power-law components,
and a fluorescence iron emission line. No
statistically significant residuals were observed in the energy range between
1 and 3 keV, and therefore no more emission lines are required to fit these
spectra. The photon index and the iron emission line were compatible within the
associated errors, but the photoelectric absorption were significantly different.}

\begin{figure*}[h!t]
  \centering
  \includegraphics[angle=-90,width=\textwidth]{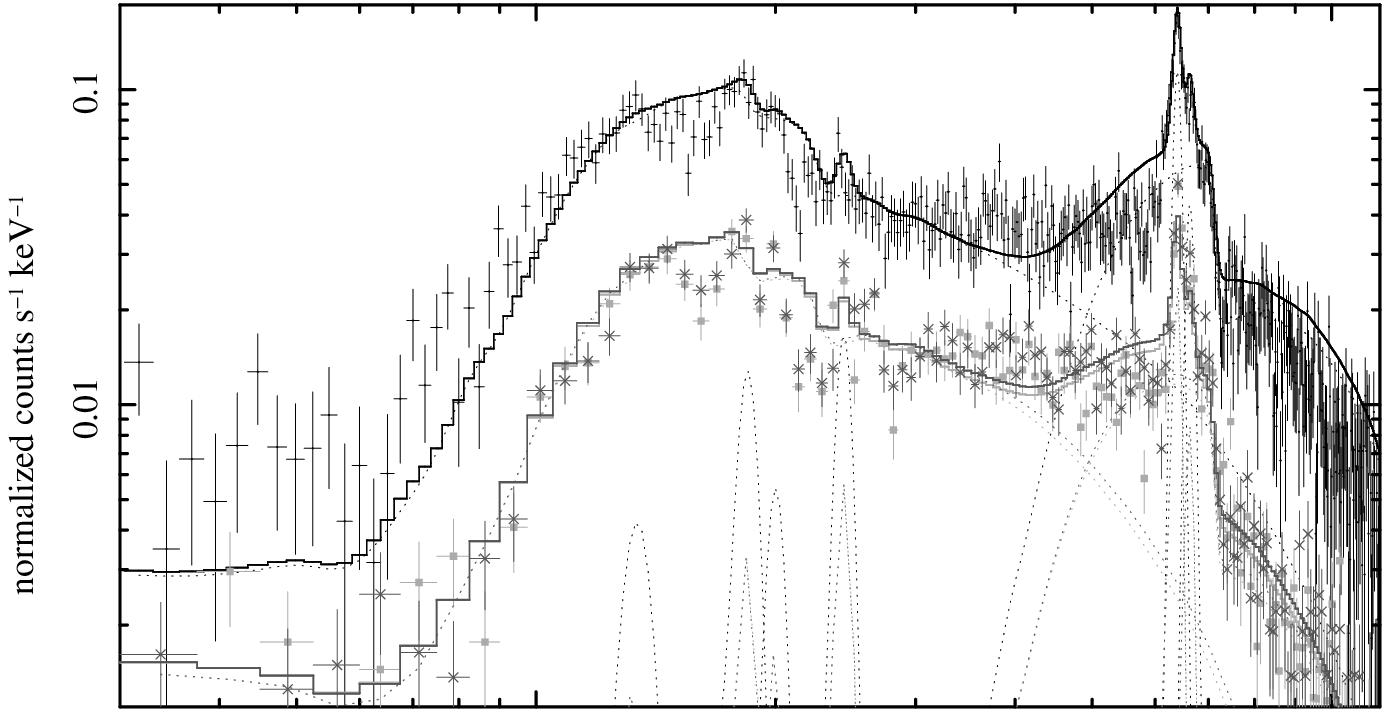}
  \includegraphics[angle=-90,width=\textwidth]{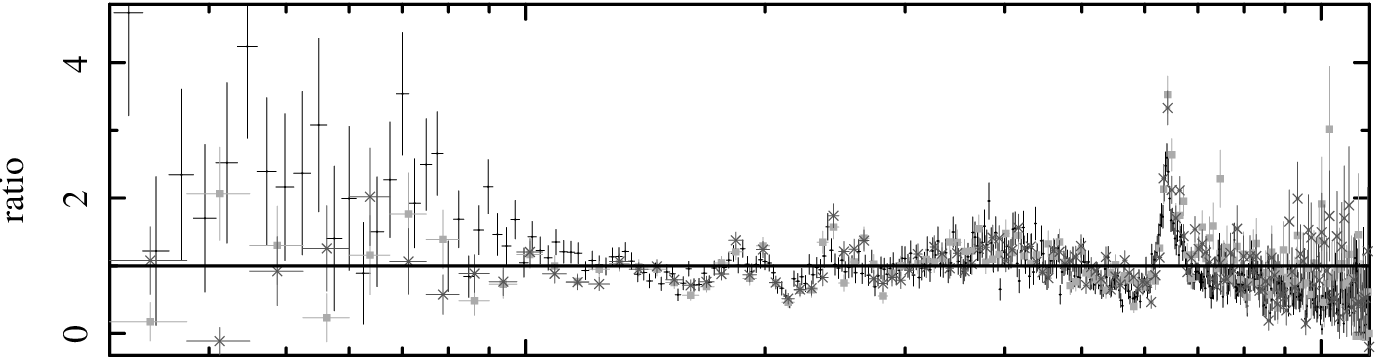}
  \includegraphics[angle=-90,width=\textwidth]{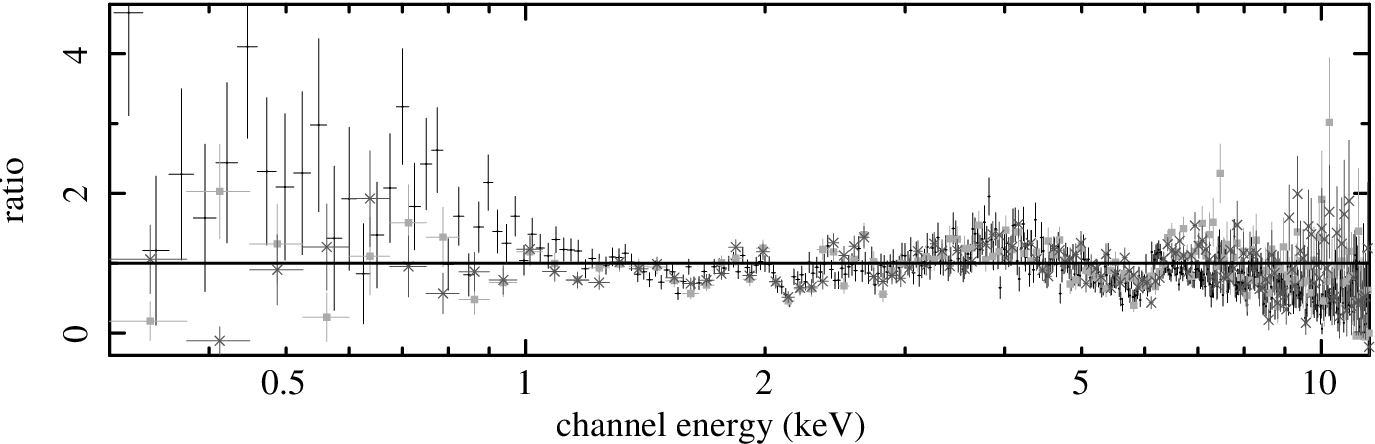}
  \caption{Iron emission lines and recombination emission lines of He- and H-like
    species in the eclipse spectrum of 4U 1538$-$52.
    \emph{Top panel}: Spectra and best fit model (two absorbed power-laws and
    six emission lines) in the 0.3--11.5 keV energy range
    obtained with \emph{PN} (top) and \emph{MOS} (light grey filled squared and dark grey cross) cameras.
    Bottom panels show the residuals.
    \emph{Second panel}: Without emission lines.
    \emph{Third panel}: With iron and recombination of He- and H-like
    emission lines for the combined spectra.
    }
  \label{eclemlines}
\end{figure*}

\textbf{The continuum spectrum in eclipse was described by two absorbed power-laws
due to the fact that the direct component should not be seen in the eclipse. In Fig.~\ref{eclemlines}
we show the eclipse spectra (top panel) and the residuals to the two absorbed power-laws
continuum model from \emph{PN} and \emph{MOS} cameras simultaneously (second panel).
Although the fit
quality is very good, the high value deduced for the absorption column is at
odds with the interpretation of this power law as the scattered component. In
fact, the values are more consistent with those of the hard component. However,
any other combination we tried, resulted in significantly worse fits. Whatever
the interpretation might be, this component seems to be the dominant one both in
and out of eclipse.}

4U 1538$-$52 shows some emission lines in the eclipse, when the X-ray
continuum emission is at a minimum. Evident are the presence of the Fe
K$\alpha$ line at 6.4 keV, an absorption edge around 7.1 keV
and emission lines between
1 and 3 keV. \textbf{We fixed the two absorbed
power-law components and added Gaussian profiles to fit these emission lines.
Table~\ref{eclcont} lists the parameters of the
continuum model we used to fit the spectra.}

\begin{table}
\begin{minipage}[t]{\columnwidth}
  \caption{Fitted parameters for the X-ray continuum of the
  eclipse spectra in Fig.~\ref{eclemlines} using \emph{PN} and \emph{MOS}
  data simultaneously.
  }
\label{eclcont}
\centering                          
\renewcommand{\footnoterule}{}      
\begin{tabular}{r r c}        
\hline\hline                 
Component & Parameter & 2PL \\    
\hline                        
{\rm Scattered}  & $N_H$ (10$^{22}$ cm$^{-2}$) & 79 $\pm$ 3 \\
  & Normalization\footnote{($\times$ 10$^{-4}$) in units of photons s$^{-1}$ cm$^{-2}$ keV$^{-1}$ at 1 keV} & 84$^{+8}_{-10}$ \\
  &  & \\
{\rm  Soft} & $N_H$ (10$^{22}$ cm$^{-2}$) & 1.08$^{+0.06}_{-0.05}$ \\
  & Normalization$^a$ & 4.17$^{+0.13}_{-0.3}$ \\
  &  & \\
  & Photon index\footnote{Both power-laws with the same photon index} & 1.74$^{+0.04}_{-0.03}$ \\
 &  C-statistic(dof) & 6452(3727) \\ 
\hline                                   
\end{tabular}
\end{minipage}
\end{table}

\textbf{Fig.~\ref{eclemlines} shows the spectra and the residuals
with respect to X-ray continuum model in the whole
energy range 0.3--11.5 keV using the \emph{PN} and \emph{MOS} cameras.
We note that in the 6.4--7.2 keV energy band some iron
emission lines are present in the residuals. As has been shown
using high resolution \emph{Chandra/HETG} data
(\cite{BS2000} 2000; \cite{dai2007} 2007),
the Fe complex can be clearly resolved into the near neutral Fe K$\alpha$
line plus hot lines from highly ionized species of Fe \textsc{XXV} and
Fe \textsc{XXVI}. Therefore,
we used Gaussian lines to fit these residuals.
In the bottom panels of Fig.~\ref{eclemlines} we show the ratio
before and after adding the emission lines described in this
Section.
}
In Table~\ref{emlieclparam},
we report the parameters of the
iron emission lines. The parameters of the Gaussian emission lines are
Energy, $\sigma$ and EW, indicating the centroid, the width and the
equivalent width, respectively.

\begin{table}
\begin{minipage}[t]{\columnwidth}
  \caption{Fitted parameters for the iron emission lines and for the
  recombination emission lines of He- and H-like species detected
  in the eclipse spectra (Fig.~\ref{eclemlines}).
    }
\label{emlieclparam}
\centering                          
\renewcommand{\footnoterule}{}      
\begin{tabular}{r r c}        
\hline\hline                 
Component & Parameter & 2PL \\    
\hline                        
{\rm Fluorescence} & Energy (keV) & 6.4050$^{+0.005}_{-0.0010}$ \\
{\rm iron line} & $\sigma$ (eV) & $\leq$ 10 \\
 & EW (eV) & 262$^{+21}_{-15}$ \\
{\rm Identification} & & Fe K$\alpha$ \textsc{I--XVII} \\
 &  & \\
{\rm Recombination} & Energy (keV) & 6.634$^{+0.020}_{-0.010}$ \\
{\rm emission lines} & $\sigma$ (eV) & $\leq$ 49 \\
  & EW (eV) & 84$^{+12}_{-14}$ \\
{\rm Identification} & & Fe \textsc{XXV} \\
 &  & \\
 & Energy (keV) & 2.4427$^{+0.0023}_{-0.023}$ \\
 & $\sigma$ (eV) & $\leq$ 18 \\
 & EW (eV) & 47$^{+15}_{-14}$ \\
{\rm Identification} & & S \textsc{XV} He$\alpha$ \\
 &  & \\
 & Energy (keV) & 2.001$^{+0.022}_{-0.05}$ \\
 & $\sigma$ (eV) & $\leq$ 280 \\
 & EW (eV) & 8$^{+14}_{-8}$ \\
{\rm Identification} & & Si \textsc{XIV} Ly$\alpha$ \\
 &  & \\
 & Energy (keV) & 1.848$^{+0.012}_{-0.024}$ \\
 & $\sigma$ (eV) & $\leq$ 103 \\
  & EW (eV) & 21$^{+17}_{-11}$ \\
{\rm Identification} & & Si \textsc{XIII} He$\alpha$ \\
 &  & \\
 & Energy (keV) & 1.34 (fixed) \\
 & $\sigma$ (eV) & 30 (fixed) \\
 & EW (eV) & $\leq$ 14 \\
{\rm Identification} & & Mg K$\alpha$, Mg \textsc{XI} He$\alpha$ \\
 &  & \\
 & C-statistic(dof) & 5148(3373) \\ 
\hline                                   
\end{tabular}
\end{minipage}
\end{table}

The EW of the neutral iron K$\alpha$ line is noticeably larger in
eclipse (262 eV) than out of eclipse ($\sim$30 eV). \cite{jjrrXrU05} found that
the EW was larger when the source flux was low by using an \emph{RXTE}
observation which covers nearly a complete orbital period.
The measured line energy
is consistent with an ionization stage up to Fe \textsc{XVII},
and this requires the ionization parameter $\xi$ should be
less than some hundreds (\cite{KMcC82}; \cite{ebisawa96}).

On the other hand, the simultaneous presence of the Fe \textsc{XXV}
line in the spectrum
implies an ionization parameter for the photoionized plasma of
$\xi \sim 10^{3.2}$ erg cm s$^{-1}$ (\cite{ebisawa96}).
\textbf{The broad range of the ionization parameter suggests either
that the emitting material is present over a wide range of distances
from the neutron star or has a large range of densities.
}

Fig.~\ref{eclemlines} also shows the residuals with respect
to a two absorbed power-law continuum model in the 1.0--3.0 keV energy
range.  Discrete recombination lines from He- and H-like
species are present in the \emph{MOS} and \emph{PN} spectra of 4U 1538$-$52.
The energy resolution of the instruments are not sufficient to
resolve the lines clearly (\cite{drake88}; \cite{hojnacki07}).
Therefore, our identification could be a blend of other emission lines.
We also fit these emission lines as Gaussians and list them in
Table~\ref{emlieclparam}.

\section{Summary and discussion}
\label{conclusion}

We presented the spectral analysis of the HMXB 4U 1538$-$52 using
an \emph{XMM-Newton} observation. The X-ray continuum is well fitted
by three absorbed power-laws with a photon index $\sim$1.13, describing
the hard, scattered and soft excess, respectively. The inferred unabsorbed
flux is $\sim 2.1 \times 10^{-10}$ erg s$^{-1}$ cm$^{-2}$ in the 0.3--11.5
keV energy band, corresponding to a luminosity of $\sim 7.5 \times 10^{35}$
erg s$^{-1}$, assuming an isotropic emission and a distance to the source of
5.5 kpc (\cite{becker}). The flux found by \emph{RXTE} in the
3--11.5 keV energy band and an orbital phase of 0.85 was
1.3$\times$10$^{-10}$ erg s$^{-1}$ cm$^{-2}$.
Using the spectrum obtained by \emph{XMM-Newton} over the same range
of phases, the flux
we obtained in this work was two times lower,
6.3$\times$10$^{-11}$ erg s$^{-1}$ cm$^{-2}$.

The soft excess is present in the spectrum and can be modelled with different
absorbed components. We simply
modelled the soft emission with an absorbed power law component, although
Fig.~\ref{PNMOS} still showed spectral residuals at lowest energies.
\textbf{Our results show that a
blackbody component could also be the physical origin of this soft excess,
taking the associated errors into account}.The soft excess in other HMXBs
has been explained by a blend of Gaussian emission lines only (\cite{boroson03}).
We also tried to fit the soft emission using Gaussian profiles, but
we did not obtain a significant improvement of the fit because
of the low level of counts below 0.6 keV.

We detected an iron K$\alpha$ line at $\sim$6.41 keV, with an EW of
$\sim$50 eV. The \emph{BeppoSAX} observation of this system obtained an EW
of 57 eV in the same orbital phase range 0.75--1.00 (\cite{robba}) and showed
an increase in the post-egress phase to 85 eV. The phase-averaged spectrum obtained
by \emph{RXTE} reported an EW of 62 eV (\cite{coburn2}). In addition this iron
line is detected in all orbital phases (\cite{robba}; \cite{rodesPhD}),
therefore the K$\alpha$ iron line is not only produced by fluorescence from less ionized iron near the neutron star's surface but also a fraction of the observed
line flux must originate from more extended regions.
\textbf{We have also detected a number of
emission lines which we interpret as recombination lines from highly ionized
species. Since these lines are detected in eclipse, they must be produced in an
extended halo. Likewise, we have found an absorption feature at 2.1 keV.
Whether it is produced by physical properties of the source or it is due to
calibration effects, is still an open issue.
}

We compared the phase-average spectrum to the eclipse spectrum.
In the phase-average spectrum, we found no evidence of any other iron line
apart from that at 6.4 keV, and the absorption
edge at $\sim$7.1 keV was well described by the X-ray absorption model.
The 6.4--7.2 keV energy band showed
a complex structure, but we did not find a proper model to describe it
or detect other iron features significantly.
We also detected discrete recombination lines in both \emph{EPIC/PN}
and \emph{EPIC/MOS} spectra. The emission lines reported in Sect.~\ref{fluorescence}
can be identified with He- and H-like ions. Although we could identify some of
them, the energy resolution of the \emph{EPIC} instruments is such that the lines
could be blended.

However, in the eclipse spectrum, we detected the fluorescence iron line
at 6.4 keV and one more iron emission line at 6.6 keV.
The discrete recombination lines detected are listed in Table~\ref{emlieclparam}.
The presence of these lines in an eclipse spectrum implies that the formation region extends
beyond the size of the B supergiant. Moreover, the ionization state was
estimated to range from $10^{2.1}$ to $10^{3.2}$ erg cm s$^{-1}$,
due to the simultaneous detection of elements with both low and high ionization levels.
\textbf{This broad range of $\xi$ also suggests either that the emitting
material is present over a wide range of distances from the
compact object or has a large range of densities.
}

\textbf{The pulse phase-resolved spectroscopy showed significant variability
of the photon index and the unabsorbed flux, but no clear correlation or
anticorrelation between them. Significant variations with the pulse phase
were also observed in the different column absorption values,
but again did not show a clear relationship to the other parameters.
}

\textbf{Future
observations with high spectral resolution instruments will be needed to
unambiguously resolve possible blended lines found in this study allowing
the full use of their diagnostic capability.}

\begin{acknowledgements}
  Part of this work was supported by the Spanish Ministry of Education
  and Science \emph{Primera ciencia con el GTC: La astronom\'{\i}a
    espa\~nola en vanguardia de la astronom\'{\i}a europea} CSD200670
  and \emph{Multiplicidad y evoluci\'on de estrellas masivas} project
  number AYA200806166C0303. This research has made use of data
  obtained through the XMM-Newton Science Archive (XSA),
  provided by European Space Agency (ESA).
  We would like to thank
  the \emph{XMM helpdesk}, particularly Matteo Guainazzi, for
  invaluable assistance in determining the systematic
  uncertainties in the \emph{PN} data.
  KLP and JPO acknowledge support from STFC.
  JMT acknowledges the support by the
  Spanish Ministerio de Educaci\'on y Ciencia (MEC) under grant
  PR2007-0176. JJRR acknowledges the support by the
  Spanish MEC under grant PR2009-0455.
\end{acknowledgements}


\begin{thebibliography}{}

   \bibitem[Arnaud 1996]{arnaud} Arnaud, K. A. 1996, in Astronomical Data Analysis Software and
   Systems V, ed. J. H. Jacoby \& J. Barnes, ASP Conf. Ser. 101, San Francisco, 17

   \bibitem[Audley et al. 1996]{AKB96} Audley, M. D., Kelley, R. L., Boldt, E. A. et al. 1996,
   ApJ, 457, 397
   
   \bibitem[Becker et al. 1977]{becker} Becker, R. H., Swank, J. H., Boldt, E. A. et al. 1977,
   ApJ, 216, L11
   
   \bibitem[Brandt \& Schulz]{BS2000} Brandt, W. N. \& Schulz, N. S. 2000, ApJ, 544, L123

   \bibitem[Boroson et al. 2003]{boroson03} Boroson, B., Vrtilek, S. D., Kallman, T. et al. 2003,
   ApJ, 592, 516
   
   \bibitem[Clark et al. 1990]{clark90} Clark, G. W., Woo, J. W., Nagase, F. et al. 1990,
   ApJ, 353, 274

   \bibitem[Clark et al. 1994]{clark94} Clark, G. W., Woo, J. W. \& Nagase, F. 1994, ApJ, 422, 336

   \bibitem[Clark 2000]{clark00} Clark, G. W. 2000, ApJ, 542, L131

   \bibitem[Coburn et al. 2002]{coburn2} Coburn, W., Heindl, W. A., Rothschild, R. E.,
   et al. 2002, ApJ, 580, 394

   \bibitem[1993]{corbet} Corbet, R. H. D., Woo, J. W. \& Nagase, F. 1993,
      A\&A, 276, 52

   \bibitem[D'A\'{\i} et al.]{dai2007} D'A\'{\i}, A., Iaria, R., Di Salvo, T. et al.
   2007, ApJ, 671, 2006

   \bibitem[Davison et al. 1977]{DWP77} Davison, P. J. N., Watson, M. G.
   \& Pye, J. P. 1977, MNRAS, 181, 73P

   \bibitem[Drake 1988]{drake88} Drake, G. W. 1988, Canadian J. Phys., 66, 586

   \bibitem[Ebisawa et al. 1996]{ebisawa96} Ebisawa, K., Day, C. S. R.,
   Kallman, T. R. et al. 1996, PASJ, 48, 425
   
   \bibitem[Elsner \& Lamb 1977]{EL1977} Elsner, R. F. \& Lamb, F. K. 1977, ApJ, 215, 897

   \bibitem[Giacconi et al. 1974]{giacconi74} Giacconi, R., Gursky, H., Kellog, E.
   et al. 1974, ApJ Suppl., 27, 37

   \bibitem[Harding \& Daugherty]{harding} Harding, A. K. \& Daugherty, J. K. 1991,
   ApJ, 374, 687.
   
   \bibitem[Hatchett \& McCray 1977]{HM77} Hatchett, S. \& McCray, R. 1977,
   ApJ, 211, 552
   
   \bibitem[Hickox et al. 2004]{HNK04} Hickox, R. C., Narayan, R. \& Kallman, T. R. 2004,
   ApJ, 614, 881
   
   \bibitem[Hojnacki et al. 2007]{hojnacki07} Hojnacki, S. M., Kastner, J. H.,
   Micela, G. et al. 2007, ApJ, 659, 585

   \bibitem[Iaria et al. 2005]{iaria05} Iaria, R., Di Salvo, T., Robba, N. R.
   et al 2005, ApJ, 634, L161

   \bibitem[Kallman \& McCray 1982]{KMcC82} Kallman, T. R. \& McCray, R. 1982, ApJS, 50, 263
   
   \bibitem[2004]{ingophd} Kreykenbohm, I. 2004, Ph. D. thesis, University of T\"ubingen

   \bibitem[Makishima et al.]{makishima87} Makishima, K., Koyama, K., Hayakawa, S.
   et al. 1987, ApJ, 314, 619
  
   \bibitem[Morrison \& McCammon 1983]{MMcC83} Morrison, R. \& McCammon, D. 1983,
   ApJ, 270, 119
   
   \bibitem[Mukherjee et al.]{mukherjee06} Mukherjee, U., Raichur, H.,
   Paul, B. et al. 2006, JAA, 27, 411

   \bibitem[Nagase et al.]{nagase92} Nagase, F., Corbet, R. H. D., Day, C. S. R. et al.
   1992, ApJ, 396, 147
   
   \bibitem[Protassov \& van Dik 2002]{ftest} Protassov, R. \& van Dik, D. A. 2002, ApJ, 571, 545
   
   \bibitem[Reynolds et al. 1992]{reynolds92} Reynolds, A. P., Bell, S. A. \& Hilditch, R. W.
   1992, MNRAS, 256, 631

   \bibitem[Robba et al. 1992]{robba92} Robba, N. R., Cusumano, G., Orlandini, M. et al.
   1992, ApJ, 401, 685
   
   \bibitem[Robba et al. 2001]{robba} Robba, N. R., Burderi, L., Di Salvo, T. et al.
   2001, ApJ, 526, 950
   
   \bibitem[Rodes 2007]{rodesPhD} Rodes, J. J. 2007, Ph.D. thesis, University of Alicante,
   http://hdl.handle.net/10045/13227
   
   \bibitem[Rodes et al. (2006)]{jjrrXrU05} Rodes, J. J., Torrej\'on, J. M. \& Bernab\'eu, G.
   2006, Proceedings of the The X-ray Universe 2005, 26-30 September 2005,
   El Escorial, Madrid, Spain. Editor: A. Wilson (Noordwijk: ESA Publications Division),
   ESA SP-604, 1, 287
   
   \bibitem[Rodes et al. 2008]{rodes2008} Rodes, J. J., Torrej\'on, J. M. \& Bernab\'eu, G.
   2006, The X-ray Universe 2008, 27-30 May 2008, Granada, Spain,
   http://xmm.esac.esa.int/external/xmm\_science/workshops/2008symposium/\#topicB

   \bibitem[Rodes-Roca et al. 2009]{jjrr09} Rodes-Roca, J. J., Torrej\'on, J. M.,
   Kreykenbohm, I. et al. 2009, A\&A, 508, 395

   \bibitem[Sako et al. 1999]{sako99} Sako M., Liedahl D. A.,
   Kahn S. M. et al. 1999, ApJ, 525, 921
   
   \bibitem[Str\"uder et al. 2001]{struder} Str\"uder, L., Briel U.,
   Dennerl, K. et al. 2001, A\&A, 365, L18
   

   \bibitem[Turner et al. 2001]{turner} Turner, M. J. L., Abbey, A., Arnaud, M. et al. 2001,
   A\&A, 365, L27
   
   \bibitem[van der Meer et al.]{vdMeer05} van der Meer, A., Kaper, L., 
   Di Salvo, T. et al. 2005, A\&A, 432, 999
   
   \bibitem[van Loon et al. 2001]{vLoon01} van Loon, J. Th., Kaper, L. \& 
   Hammerschlag-Hensberge, G. 2001, A\&A, 375, 498
   
   \bibitem[White et al. 1983]{WSH83} White, N. E., Swank, J. H. \& Holt, S. S. 1983,
   ApJ, 270, 711
   
   
\end{thebibliography}
\end{document}